\documentclass[12pt]{JHEP3}
\usepackage{graphicx}
\usepackage{dcolumn}
\usepackage{bm}
\usepackage{amsmath,amssymb,amscd}
\usepackage{longtable}

\usepackage{caption}

\newcommand{\be}{\begin{eqnarray}}
\newcommand{\ee}{\end{eqnarray}}

\newcommand{\bn}{\begin{enumerate}}
\newcommand{\en}{\end{enumerate}}

\newcommand{\beqn}{\begin{eqnarray}}
\newcommand{\eeqn}{\end{eqnarray}}

\def\IZ{\mathbb{Z}}

\def\CL{{\cal L}}

\def\Tr{{\rm Tr}}
\def\tr{{\rm tr}}

\newcommand{\ket}[1]{\vert  #1\rangle}

\newcommand{\RR}{{\mathbb R}}

\newcommand{\cN}{{\mathcal N}}

\newcommand{\Hom}{{\rm Hom}}
\newcommand{\cW}{{\mathcal W}}

\title{Seiberg-like Dualities for 3d ${\cN}=2$  Theories with $SU(N)$ Gauge Group}

\author{ Jaemo Park$^{1,2}$, Kyung-Jae Park$^{1}$

\\

$^1$Department of Physics, POSTECH, Pohang 790-784, Korea
\\
$^2$Postech Center for Theoretical Physics (PCTP), Postech, Pohang
  790-784, Korea

\\
\\
E-mail: \email{ jaemo@postech.ac.kr, jaco@postech.ac.kr} } 

\abstract{We work out Seiberg-like dualities for 3d $\cN=2$ theories with $SU(N)$ gauge group.
We use  the $SL(2,\IZ)$ action on 3d conformal field theories with $U(1)$ global symmetry. One of generator
S of $SL(2,\IZ)$
 acts as gauging of the $U(1)$ global symmetry.  Utilizing $S=S^{-1}$ up to charge conjugation,
we obtain Seiberg-like dual of $SU(N)$ theories by gauging topological U(1) symmetry of the Seiberg-like dual
of $U(N)$ theories with the same matter content. We work out the Aharony dualities for $SU(N)$ gauge theory
with $N_f$ fundamental/anti-fundamnetal flavors, with/without one adjoint matter with the superpotential. We also work out
the Giveon-Kutasov dualities for $SU(N)$ gauge theory with Chern-Simons term and with $N_f$ fundamental/anti-fundamental flavors.
For all the proposed dualities, we give various evidences such as chiral ring matching and the superconformal index computations.
We find the perfect matchings.}

\begin{document}

\section{Introduction}

Recently, there has been renewed interest in nonperturbative dualities between
three dimensional theories such as mirror symmetry and Seiberg-like
dualities. This is explained in part by the availability of
sophisticated tools such as the partition function on $S^3$ and the
superconformal index. Using
these tools, one can give impressive evidence
for various 3d dualities. Some of works in this area are \cite{Giveon09}-\cite{SeibergCS}. One can also obtain
the R-charge of the fields by maximizing the free energy of the
theory of interest \cite{Jafferis10}.

In this paper we continue this line of research and study Seiberg-like dualities \cite{Giveon09, Aharony97} for $\cN=2$ $d=3$ gauge theories with
$SU(N_c)$ group. In 3-d, Seiberg-like dualities are known for $U(N_c)$ theories but the duality for
$SU(N_c)$ theories remains elusive. If we knew the dual pairs of $SU(N_c)$ theories, we could obtain
the dual pairs of $U(N_c)$ theories by gauging $SU(N_c)$ theories. On the other hand to obtain $SU(N_c)$ dual pairs out of
$U(N_c)$ dual pairs, we need ungauging overall $U(1)$ of $U(N_c)$ gauge symmetry, which is not obvious how to do it.

The purpose of this paper is to propose such ungauging procedure to obtain Seiberg-like dualities for $SU(N_c)$ theories.
The idea comes from the observation that there's $SL(2, \IZ)$ transformation for 3d conformal field theories with
$U(1)$ global symmetry \cite{Witten-SL2}. In particular, one of the generator S of $SL(2, \IZ)$ involves the gauging
of $U(1)$ global symmetry and introduces a new $U(1)$ flavor symmetry, often called 'topological' symmetry.
This is the perfect setting to gauge $SU(N_c)$ theories since we can obtain $U(N_c)$ theories by gauging the overall $U(1)$ global
symmetry of $SU(N_c)$ theories and new flavor symmetry of $U(N_c)$ theory is the topological symmetry $U(1)_T$, whose
current is given by $\ast d {\rm Tr} A$ where ${\rm Tr}A$ denotes the overall $U(1)$ gauge field of $U(N_c)$ theories.
Since $S^2=C$ with $C$ being charge conjugation, by applying gauging the $U(1)_T$ we can obtain the same superconformal
field theories as the original $SU(N_c)$ theories up to charge conjugation. With respect to $U(1)_T$,
 monopole operators are charged, which has the nonperturbative origin in the original theory. Thus we had better carry out
 the 2nd $S$-operation or gauging $U(1)_T$ in the Seiberg-like dual pair of $U(N_c)$ theory. If the matter contents are
 charge conjugation invariant, we obtain Seiberg-like dual pair of the original $SU(N_c)$ theory by this way.
 So we can obtain the Seiberg-like dual of any $SU(N_c)$ theory, if the Seiberg-like dualities are known for
 $U(N_c)$ theory with the same matter contents as the $SU(N_c)$ theory. All we have to is to gauge $U(1)_T$ of Seiberg-like
 dual pair of the given $U(N_c)$ theory.
  We apply this idea to various $SU(N_c)$ theories to obtain Aharony duals \cite{Aharony97}.  Giveon-Kutasov dualities \cite{Giveon09} for $SU(N_c)$ gauge theories with Chern-Simons term can be obtained from Aharony dualities by giving
  axial mass to some of flavors.
 We subject such candidates of dual pairs to various tests such as the computation of the superconformal index to find
 the perfect matching.

 The content of the paper is as follows;
 In section 2, we review the basics of the superconformal index in 3-dimensions.
 In section 3, we review the $SL(2, \IZ)$ transformation of 3d conformal field theories with $U(1)$ flavor symmetry
  and emphasize S-operation in relation to gauging $U(1)$ flavor symmetry.
  In section 4, we consider the simplest example of $U(1)$ gauge theory with one flavor and carry out gauging/ungauging
  operation.
 In section 5, we work out Aharony dualities of $SU(N_c)$  gauge theories with $N_f$ fundamental and anti-fundamental flavors
 and carry out various tests to find the nice agreements.
 In section 6, we obtain Giveon-Kutasov dualities for $SU(N_c)$ gauge theories with $N_f$ fundamental and anti-fundamental flavors and with Chern-Simons term. These dualities are obtained from Aharony dualities of the section 5 by giving axial mass to
 some flavors.
 In section 7, we obtain Aharony dual of $SU(N_c)$ gauge theories with $N_f$ fundamental and anti-fundamental flavors,
 one adjoint $X$ and with the superpotential $W= {\rm Tr} X^{n+1}$. Again we put the proposal to various tests and find the
 perfect matching.

As this work is finished, we receive the paper \cite{Seibergsu} which has overlap with ours.

\section{3d superconformal index}

Let us discuss the superconformal index for  $\cN=2$ $d=3$
superconformal field theories (SCFT). The bosonic subgroup of the 3d
$\cN=2$ superconformal group  is $SO(2,3) \times SO(2) $. There are
three Cartan elements denoted by $\epsilon, j_3$ and $R$ which come
from three factors $SO(2)_\epsilon \times SO(3)_{j_3}\times SO(2)_R
$ in the bosonoic subgroup, respectively. The superconformal index
for an $\cN=2$ $d=3$ SCFT is defined as follows
\cite{Bhattacharya09}:
\begin{equation}
I(x,t)=\Tr (-1)^F \exp (-\beta'\{Q, S\}) x^{\epsilon+j_3}\prod_a
t_a^{F_a} \label{def:index}
\end{equation}
where $Q$ is a   supercharge with quantum numbers $\epsilon =
\frac{1}2, j_3 = -\frac{1}{2}$ and $R=1$, and $S= Q^\dagger$.  The
trace is taken over the Hilbert space in the SCFT on
$\mathbb{R}\times S^2$ (or equivalently over the space of local
gauge-invariant operators on $\RR^3$). The operators $S$ and $Q$
satisfy the following anti-commutation relation:
\begin{equation}
 \{Q, S\}=\epsilon-R-j_3 : = \Delta.
\end{equation}
As usual, only BPS states satisfying the bound $\Delta =0 $
contribute to the index, and therefore the index is independent of
the parameter $\beta'$. If we have additional conserved charges
$F_a$ commuting with the chosen supercharges ($Q,S$), we can turn on
the associated chemical potentials $t_a$, and then the index counts
the number of BPS states weighted by their quantum
numbers.

The superconformal index is exactly calculable using the
localization technique \cite{Kim09,Imamura11}.  It can be written in
the following form:
\begin{eqnarray}\label{index32}
& &I(x,t)= \nonumber\\
& &\sum_{m\in \mathbb{Z}} \int da\, \frac{1}{|\cW_m|}
e^{-S^{(0)}_{CS}(a,m)}e^{ib_0(a,m)} \prod_a t_{a}^{q_{0a}(m)}
x^{\epsilon_0(m)}\exp\left[\sum^\infty_{n=1}\frac{1}{n}f_{tot}(e^{ina},
t^n,x^n)\right].\qquad
\end{eqnarray}

The origin of this formula is as follows.
 To compute the trace over the Hilbert space on $S^2\times\RR$, we use path-integral on $S^2\times S^1$ with
 suitable boundary conditions
on the fields. The path-integral is evaluated using localization,
which means that we have to sum or integrate over all BPS saddle
points. The saddle points are spherically symmetric configurations
on $S^2\times S^1$ which are labeled by magnetic fluxes on $S^2$ and
holonomy along $S^1$. The magnetic fluxes are denoted by  $\{m\}$
and take values in the cocharacter lattice of $G$ (i.e. in
$\Hom(U(1),T)$, where $T$ is the maximal torus of $G$), while the
eigenvalues of the holonomy are denoted $\{ a \}$ and take values in
$T$. $S_{CS}^{(0)}(a,m)$ is the classical action for the
(monopole+holonomy) configuration on $S^2\times S^1$,
$\epsilon_0(m)$ is the Casmir energy of the vacuum state on $S^2$
with magnetic flux $m$, $q_{0a}(m)$ is the $F_a$-charge of the
vacuum state, and $b_0(a,m)$ represents the contribution coming from
the electric charge of the vacuum state. The last factor comes from
taking the trace over a Fock space built on a particular vacuum
state. $|\cW_m|$ is the order of the Weyl group of the part of $G$
which is left unbroken by the magnetic fluxes $m$ . These
ingredients in the formula for the index are given by the following
explicit expressions:
\begin{eqnarray}\label{components}
&&S^{(0)}_{CS}(a,m) = i \sum_{\rho\in R_{F}} k \rho(m) \rho(a) , \\
&&b_0(a,m)=-\frac{1}{2}\sum_\Phi\sum_{\rho\in R_\Phi}|\rho(m)|\rho(a),\nonumber\\
&&q_{0a}(m) = -\frac{1}{2} \sum_\Phi \sum_{\rho\in R_\Phi} |\rho(m)| F_a (\Phi), \nonumber \\
&& \epsilon_0(m) = \frac{1}{2} \sum_\Phi (1-\Delta_\Phi)
\sum_{\rho\in R_\Phi} |\rho(m)| - \frac{1}{2} \sum_{\alpha \in G}
|\alpha(m)|, \nonumber
\end{eqnarray}
\begin{eqnarray}
&& f_{tot}(x,t,e^{ia})=f_{vector}(x,e^{ia})+f_{chiral}(x,t,e^{ia}),\nonumber\\
&& f_{vector}(x,e^{ia})=-\sum_{\alpha\in G} e^{i\alpha(a)} x^{|\alpha(m)|},\nonumber \\
&& f_{chiral}(x,t,e^{ia}) = \sum_\Phi \sum_{\rho\in R_\Phi} \left[
e^{i\rho(a)} \prod_a t_{a}^{F_{a}}
\frac{x^{|\rho(m)|+\Delta_\Phi}}{1-x^2}  -  e^{-i\rho(a)} \prod_a
t_{a}^{-F_{a}} \frac{x^{|\rho(m)|+2-\Delta_\Phi}}{1-x^2}\nonumber
\right]\label{universal}
\end{eqnarray}
where $\sum_{\rho\in R_F}, \sum_\Phi$, $\sum_{\rho\in R_\Phi}$ and
$\sum_{\alpha\in G}$ represent summations over all fundamental
weights of $G$, all chiral multiplets, all weights of the
representation $R_\Phi$, and all roots of $G$, respectively.

We will find it convenient to rewrite the integrand in \eqref{index32}
as a product of contributions from the different multiplets.  First,
note that the single particle index $f$ enters via the so-called
Plethystic exponential:
\begin{align}  P. E. (f(x,t,z=e^{ia}, m)\equiv \exp \bigg( \sum_{n=1}^\infty \frac{1}{n} f ( x^n, {t}^n,{z}^n=e^{ina}, m) \bigg) \end{align}
while we define $z_j=e^{ia_j}$.
Specifically, consider a single chiral field $\Phi$, whose single
particle index is given by\footnote{Note that $a$ in $\rho(a)$ and
the subscript $a$ in $t_a$ or $f_a$ denotes the different object.} :
\begin{align} \sum_{\rho \in R_\Phi} \bigg( e^{i \rho(a)} {t_a}^{f_a(\Phi)} \frac{x^{ |\rho(m)|
+ \Delta_\Phi}}{1-x^2} - e^{-i \rho(a)} {t_a}^{-f_a(\Phi)} \frac{x^{
|\rho(m)| + 2-\Delta_\Phi}}{1-x^2} \bigg). \end{align}
Then the index contribution is simply
\begin{equation}
\prod_{\rho \in R_\Phi} P.E. \bigg( e^{i \rho(a)} {t_a}^{f_a(\Phi)} \frac{x^{ |\rho(m)|
+ \Delta_\Phi}}{1-x^2} - e^{-i \rho(a)} {t_a}^{-f_a(\Phi)} \frac{x^{
|\rho(m)| + 2-\Delta_\Phi}}{1-x^2} \bigg)
\end{equation}

The full index will involve a product of such factors over all the
chiral fields in the theory, as well as the contribution from the
gauge multiplet.  It is given by:
\begin{align} I(x,t) = \sum_{m\in\mathbb{Z}} \oint \prod_j \frac{dz_j}{2 \pi i z_j} \frac{1}{|\mathcal W_m|} e^{-S_{CS}(m,a)} Z_{gauge}(x,z,m)
\prod_\Phi Z_\Phi(x,t,z,m) \end{align} where:
\begin{align*} Z_{gauge}(x,z=e^{ia},m) = \prod_{\alpha \in ad(G)} x^{-|\alpha(m)| } \bigg(1 - e^{i \alpha(a)}
x^{2 |\alpha(m)|} \bigg),  \end{align*}
\begin{eqnarray}
 Z_\Phi(x,t,z,m)
 &=& \prod_{\rho \in R_\Phi}  \!\!\bigg(\!\!x^{(1- \Delta_\Phi) }
e^{-i \rho(a)} \prod_a {t_a}^{-f_a(\Phi)} \!\!\bigg)^{|\rho(m)|/2}  \nonumber \\
& & P.E. \bigg( e^{i \rho(a)} {t_a}^{f_a(\Phi)} \frac{x^{ |\rho(m)|
+ \Delta_\Phi}}{1-x^2} - e^{-i \rho(a)} {t_a}^{-f_a(\Phi)} \frac{x^{
|\rho(m)| + 2-\Delta_\Phi}}{1-x^2} \bigg)
\end{eqnarray}
Note that by shifting $t_a\rightarrow t_ax^{c_a}$, one can change
the value of the R-charge $\Delta_{\Phi}$. Hence $\Delta_{\Phi}$
remains the free parameter for generic cases.

The above is the ordinary superconformal index.  We need  two more generalizations
 for later purposes. The first one is the notion of the generalized
index. When we turn on the chemical potential $t_a$, which can be
regarded as turning on a Wilson line for a fixed background gauge
field. The generalized index is obtained when we turn on the
nontrivial magnetic flux $n_a$ for the corresponding background
gauge field. Only the contribution to the chiral multiplets are
changed and this is given  by the replacement $\rho(m)\rightarrow
\rho(m)+\sum_a f_a(\Phi)n_a$
\begin{eqnarray}
& &Z_\Phi(x,t,z=e^{ia},m)
 =\prod_{\rho \in R_\Phi}  \!\!\bigg(\!\!x^{(1- \Delta_\Phi) }
e^{-i \rho(a)} \prod_a {t_a}^{-f_a(\Phi)} \!\!\bigg)^{|\rho(m)|/2+\sum_a f_a(\Phi)n_a/2}  \nonumber \\
& & P.E. \bigg( e^{i \rho(a)} {t_a}^{f_a(\Phi)} \frac{x^{ |\rho(m)+\sum_a f_a(\Phi)n_a|
+ \Delta_\Phi}}{1-x^2} - e^{-i \rho(a)} {t_a}^{-f_a(\Phi)} \frac{x^{
|\rho(m)+\sum_a f_a(\Phi)n_a| + 2-\Delta_\Phi}}{1-x^2} \bigg) \nonumber
\end{eqnarray}
Here   $n_a$ should take integer value as does  $m_j$.

For every $U(N)$ gauge group, we can define another abelian symmetry
$U(1)_T$ whose conserved current is $*F$ of overall $U(1)$ factor.
To couple this topological current to background gauge field we
introduce $BF$ term $\int A_{BG} \wedge \tr dA+ \cdots$ and terms
needed for supersymmetric completion. This introduces to the index
\begin{equation}
z^nw^{\sum_j m_j}
\end{equation}
where $n$ is the new discrete parameter representing the topological
charge while $w$ is the chemical potential for $U(1)_T$.

\section{ $SL(2, \IZ)$ action on the 3d CFTs with $U(1)$ symmetry}\label{sl2z}
Gauging and ungauging of $U(1)$ factor we adpoted in this paper is closely related to the $S$-operation
for the 3d conformal field theories (CFTs) with $U(1)$ flavor symmetry.
It was found that there is a $SL(2,\IZ)$ action on the space of
3-dimensional conformal field theories with $U(1)$ flavor symmetry.
This action was first described in \cite{Witten-SL2} as a way to
understand the meaning of different choices of boundary conditions
for an abelian gauge field in $AdS_4$ in the context of
$AdS_4/CFT_3$. And in \cite{DGG, DGG2} such $SL(2,\IZ)$ action on 3d abelian gauge theories with $U(1)$ flavor symmetry
was considered. We closely follow their explanation in the below.

 $SL(2,\IZ)$ acts on the space of 3d theories equipped
with a specific way to couple a $U(1)$ flavor symmetry to a
background $U(1)$ gauge field. The $SL(2,\IZ)$ action has two generators $S, T$ satisfying
\begin{equation}
 S^4 = (ST)^3 = I
 \end{equation}
The  $T$-opration on the 3d conformal theories  only
modifies the prescription of how to couple the theory to the
background gauge field $A$, by adding $\ast F=\ast dA$ to the conserved current for
the background $U(1)$ symmetry. At the Lagrangian level, this means
adding a background Chern-Simons interaction at level $k=1$,
\begin{equation}
 T\,:\quad \CL\to \CL + \frac{1}{2} A\wedge dA\,.
\end{equation}

On the other hand, the $S$-operation changes the structure of the 3d theory by making the background gauge field $A$ 
dynamical.\footnote{One can add a Yang-Mills kinetic term at intermediate stages in the calculation.
But for $S$ to have the correct properties, one must flow to the IR at the end, and then $g_{{\rm YM}}\to \infty$ and this term is removed.} Once the old $U(1)$ flavor symmetry turns into gauge field, it has the new $U(1)$ flavor current
given by $\ast F$ of the gauged $U(1)$. At the Lagrangian level
\begin{equation}
 S\,:\quad \CL\to \CL+  A_{\mathrm{new}} \wedge dA
\qquad\text{($A$ dynamical)}\,. \label{s}
\end{equation}

Monopole operators for $A$ are charged under the new
$U(1)$ flavor symmetry, hence this $U(1)$ is sometimes called
topological.
Since $S^2=C$, if one carries out the gauging twice following the prescription eq. (\ref{s}) , the resulting theory is equivalent to the
original theory up to charge conjugation. One can also say that gauging $U(1)$ (S) is equivalent to
ungauging $U(1)$ up to charge conjugation ($S^{-1}C$). When we work out gauging/ungauging $U(1)$, we are always intended to apply
$S$ / $S^{-1}$ operation.

 From the definitions of $S$ and $T$, one can prove
that the relations $S^2=C$ and $(ST)^3=I$ hold, where the
transformation $C$ (charge conjugation) just inverts the sign of the
background gauge field.

This has a suitable $\cN=2$ generalization. Suppose that we have a theory with $U(1)$ global symmetry
coupled to a background vector multiplet $V$. $V$ has a real scalar $\sigma$, two Majorana fermions $\lambda^{\alpha}$
and the gauge field $A$. This can be dualized to a linear multiplet $\Sigma$
\begin{equation}
V\leftrightarrow \Sigma= D^{\alpha}D_{\alpha} V
\end{equation}
with the lowest component of $\Sigma$ being $\sigma$.
Now in order to supersymmetrize $SL(2, \IZ)$ action,
we simply have to substitute $V\Sigma'$ for $A\wedge dA'$.
In particular, $S$-operation is given by
\begin{equation}
\CL(V) \rightarrow \CL(V)+ \int d^4\theta \Sigma_{new} V
\end{equation}

Now we can apply this idea to obtain Aharony dual of $SU(N)$ gauge theory.
The basic idea is that we start from $SU(N)$ gauge theory with
matters which are invariant under the charge conjugation so that if
we apply S-operation twice we are back to the original theory. Given
$SU(N)$ gauge group we have obvious global $U(1)$ symmetry and if we apply
the S-operation we introduces the gauging of $U(1)$ theory with the
BF type coupling to the background gauge field. Thus we now have the
$U(N)$ gauge theory with the usual $U(1)_T$ topological symmetry for which
the monopole operators are charged. This is the typical setting
where Aharony duality of $U(N)$ gauge theory is discussed. If we
apply S-operation again, then we gauge topological symmetry and
introduce another $U(1)$ flavor symmetry. By this procedure we are
back to our original theory of $SU(N)$ theory assuming the matter contents is charge-conjugation
invariant. Thus gauging
 topological $U(1)$ corresponds to ungauging overall $U(1)$
gauge symmetry. On the other hand, by applying the same S-operation
to the Aharony dual theory of $U(N)$ theory, we obtain Aharony-dual
of $SU(N)$ theory.

For example, if we start from $U(N)$ gauge theory with $N_f$
fundamental flavors, Aharony dual is given by $U(N-N_f)$ gauge
theory with $N_f$ flavors with the following superpotentials
\begin{equation}
W=v_{+}V_{-}+v_{-}V_{+}+ Mq\tilde{q}
\end{equation}
where $M, v_{\pm}$ is the singlet for $U(N-N_f)$ corresponding to
the mesons and the monopole operators for $U(N)$ gauge theory. By
applying the S-operation we obtain $U(1)_T \times U(N-N_f)$ gauge
theory with the above superpotential and the monopole operators
$v_{\pm}, V_{\pm}$ is charged under $U(1)_T$. Furthermore, we have the
additional BF type coupling
\begin{equation}
A_{T} \wedge d {\rm Tr} A
\end{equation}
where $ {\rm Tr} A$ denotes the overall $U(1)$ gauge field of $U(N-N_f)$.
Following the above logic this should be the Aharony dual of $SU(N)$
with $N_f$ flavors. We subject this claim to the various tests in
the next section.

Furthermore this logic applies any $SU(N)$ gauge theory with
charge-conjugation invaraint matter contents to obtain Aharony dual if the corresponding
Aharony dual of $U(N)$ theory is known. Thus we also consider the
theory of $U(N)$ theory with $N_f$ flavors and adjoint matter $X$
with the superpotential $W={\rm Tr} X^{n+1}$ and work out its Aharony dual.
One should note that starting from $SU(N)$ theory one can generate whole classes of SCFTs
by $SL(2, \IZ)$ transformation. We are currently working out the details such SCFTs \cite{toappear}.

For later purpose we need to work out how the index would transform
under the S-transformation. Suppose that $\tilde{I}(z,s)$ denotes
the generalized index with $U(1)$ global symmetry and $z,s$ are
respectively chemical potential and magnetic flux associated with the $U(1)$
global symmetry. Let us denote the generalized index of
S-transformed theory by $I(u,m)$ with $u, m$ are respectively chemical
potential and magnetic flux associated with new $U(1)$ global
symmetry. Then $I$ is given by
\begin{equation}
I(w,m)=\sum_{s \in Z} \oint \frac{dz}{2\pi i z}w^s
z^{m}\tilde{I}(z,s) \label{eq2}
\end{equation}
The relation between the index and the S-transformed index is well
known. It's convenient to use the charge basis \cite{DGG2}
\begin{equation}
\tilde{I}(z,s)=\sum_{e \in Z}\tilde{I}(e,s)z^e, \,\,\,I(w,m)=\sum_{e
\in Z}I(e,m)w^e
\end{equation}
Then the right hand side of (\ref{eq2}) becomes
\begin{equation}
\sum_{s \in Z} \int \frac{dz}{2\pi i z}w^s
z^{m}\tilde{I}(z,s)=\sum_{s \in Z} \sum_{e \in Z}\int \frac{dz}{2\pi
i z}w^s z^{m+e}\tilde{I}(e,s)=\sum_{e \in Z}w^s\tilde{I}(-m,s)
\end{equation}
which is equal to $I(w,m)=\sum_{e \in Z}I(e,m)w^e$. Thus we have
\begin{equation}
I(e,m)=\tilde{I}(-m,e).
\end{equation}
Note that in the charge basis S-operation takes the simple form
\begin{equation}
\left( \begin{array}{c}
                e \\
                m      \end{array} \right) \rightarrow \left( \begin{array}{cc}
                0 & 1 \\
               -1 & 0      \end{array}  \right) \left( \begin{array}{c}
                e \\
                m      \end{array} \right)
\end{equation}
Thus we have
\begin{equation}
\tilde{I}(z,s)=\sum_{e \in Z}z^e\int \frac{dw}{2\pi i w^{s+1}}
I(w,-e). \label{ungauge}
\end{equation}
We regard this formula as ungauging $U(1)$. We also have the inverse
relation
\begin{equation}
I(u,-m)=\sum_{e \in Z}u^e\int \frac{dz}{2\pi i z^{m+1}}
\tilde{I}(z,e) \label{ungauge}
\end{equation}
Let's denote the generalized index of the theory obtained from
S-operation on the theory  with index being $I(u,m)$ by
$I'(z,s)$. One can easily check  that $\tilde{I}(z,s)=I'(z^{-1},
-s)$. Or in charge basis $\tilde{I}(e,s)=I'(-e, -s)$, which is the
consequence of $S^2=C$. If we consider $U(N)$ theory with $N_f$
flavors, the theory obtained by S-transformation differs from
$SU(N)$ theory with $N_f$ flavors by the sign of the charges of the
matter. Thus the role of chiral multiplet and the anti chiral
multiplet is exchanged. But since we are dealing with the same
number of chiral multiplets and the anti chiral multiplets, we
obtain the same theory. However when we interpret the index result,
we should keep in mind of such sign flipping.

\section{Ungauging $\mathcal N=2$ $N_f=1$ SQED and its Ahrony dual}

Let's consider the simplest example of $\mathcal N=2$ Aharony dual pair. $\mathcal N=2$
SQED with $N_f=1$ flavor. Its dual is given by XYZ model. In using
the  convention of the section \ref{sl2z}, this is the theory with no gauge group with the
superpotential
\begin{equation}
W=v_{+}v_{-} M  \label{superpo}
\end{equation}
where $v_{\pm}$ is charged under $U(1)_T$. If we ungauge
$U(1)$ gauge group for SQED, we are left with the free theory with $N_f=1$
flavor, since  this corresponds to $S^{-1}$-operation. This is the $\mathcal N=4$
theory with one free hypermultiplet. On the other hand, if we gauge
$U(1)_T$ of XYZ model, we obtain $U(1)$ theory with $N_f=1$
flavor $v_{\pm}$ with additional neutral chiralmultiplet $M$ whose
superpotential is given by eq. (\ref{superpo}). This is $\mathcal N=4$ SQED
with one hypermultiplet. Thus $\mathcal N=4$ theory with one free hypermultiplet and
$\mathcal N=4$ SQED
with one hypermultiplet are related by $S^2=C$. But since the matter content is C-invariant, we obtain
the equivalent theory. This is nothing but the simplest mirror pair.
Thus we can regard this mirror pair as a special case of Aharony
duality of $SU(N)$ theory with $N_f$ flavors with $N=N_f=1$.

This simple example also shows that why S-operation involves the duality transformation.
Starting from $\cN=4$ one free hypermultiplet, we obtain SQED with $N_f=1$. Under the $U(1)_T$ symmetry,
monopole operators of SQED are charged. Thus in order to carry out the gauging of $U(1)_T$ we had better
go to the frame where the monopole operators are elementary fields. This is possible if we work in the Aharony dual
of SQED with $N_f=1$. This is nothing but the XYZ model and $U(1)_T$ is mapped to the usual $U(1)$ global symmetry.
Hence gauging $U(1)_T$ is straightforward and we obtain $\cN=4$ SQED with one hypermultiplet.
Following this example, we carry out gauging of $U(1)_T$ for the Seiberg-like dual of original $U(N_c)$ theories to
obtain Seiberg-like dual of $SU(N_c)$ theories.

\subsection{Index of Ungauged SQED $N_f=1$}
As a warmup exercise of the index gymnastics, we consider the ungauging of SQED with one flavor.
Since our major concern is the gauge symmetry and the
topological $U(1)_T$ symmetry, we will turn off the chemical potential for $U(1)$
axial symmetry. Similar manipulation will be used for handling $U(N)$ theory with $N_f$ flavors.
The index formula of the  SQED with one flavor
is given by .
\beqn
I(w,m_w;x)&=&\sum_{m_1\in \mathbb Z}\oint\frac{dz_1}{2\pi iz_1}w^{m_1}  z_1^{m_w} Z_{Q} Z_{\tilde Q}\nonumber\\
&=&\sum_{m_1\in \mathbb Z}\oint\frac{dz_1}{2\pi iz_1}w^{m_1}
z_1^{m_w} Z_{\Phi}(z_1,m_1;x) \eeqn\label{aa} where $z_1$ is the
holonomy of $U(1)_T$ and $w$ is the chemical potential for the
background gauge field coupled to $U(1)_T$.  $Z_Q$ and
$Z_{\tilde Q}$ is some function of $x$, $z_1$ and $m_1$. If we
ungauge $U(1)$, we expect to have the free theory with $N_f=1$. Using the formula eq. (\ref{ungauge}), the
ungauged index is
 \beqn
\tilde{I}(z,s;x)&=&\sum_{m_w\in \mathbb Z}z^{-m_w}\oint \frac{dw}{2\pi i w^{s+1}}I(w,m_w;x)\nonumber\\
&=&\sum_{m_w\in \mathbb Z}z^{-m_w}\oint \frac{dw}{2\pi i
w^{s+1}}\left(\sum_{m_1\in \mathbb Z}\oint\frac{dz_1}{2\pi
iz_1}w^{m_1}z_1^{m_w} Z_{\Phi}(z_1,m_1;x) \right)\label{qq1} \eeqn We
can expand the $Z_{\Phi}(z_1,m_1;x)$ in the integer power of $z_1$,
such as $Z_{\Phi}(z_1,m_1;x)=\sum_{z_1\in\mathbb Z}z_1^n \tilde
Z_{\Phi}(n,m_1;x)$. So eq. \eqref{qq1} is equal to
\beqn
\tilde{I}(z,s;x)=\sum_{m_w\in \mathbb Z}z^{-m_w}\oint \frac{dw}{2\pi
i w^{s+1}}\left(\sum_{m_1\in \mathbb Z}\oint\frac{dz_1}{2\pi
iz_1}w^{m_1}z_1^{m_w}\sum_{n\in\mathbb Z}z_1^n \tilde
Z_{\Phi}(n,m_1;x) \right). \eeqn\label{ungaugedu1}
The above
integral  has very simple dependence on $w$ and $z_1$. Integration
over $z_1$ gives simply the restriction that $n=-m_w$ and the
integration over $w$ gives $m_1=s$. So $\tilde I(z,s)$
becomes
\beqn \tilde I(z,s;x)=\sum_{m_w\in\mathbf Z}z^{-m_w}\tilde
Z_{\Phi}(-m_w,s;x)=Z_{\Phi}(z,s;x) \eeqn
Explicit form of this ungauged
index is $\tilde I(z,s;x)=Z_{Q}Z_{\tilde Q}$, where $Z_{Q}$ and
$Z_{\tilde Q}$ is \beqn
Z_{Q}=\left(x^{1-r}z^{-1}\right)^{|s|/2}P.E.\left(\frac{z x^r-z^{-1}x^{2-r}}{1-x^2}x^{|s|}\right)\nonumber\\
Z_{\tilde
Q}=\left(x^{1-r}z\right)^{|s|/2}P.E.\left(\frac{z^{-1}x^r-zx^{2-r}}{1-x^2}x^{|s|}\right).\label{qq}
\eeqn
The resulting index is that of the free theory with a chiral and an
anti-chiral field as expected. The chemical potential of the $U(1)$ global symmetry
is  $z$.

\subsection{Index of ungauged (or gauged) XYZ}
Let's do the same process to the dual XYZ theory. Then the index of ungauged XYZ
 theory becomes \beqn
\tilde{I}(z,s;x)&=&\sum_{m_w\in \mathbb Z}z^{-m_w}\oint \frac{dw}{2\pi i w^{s+1}}I(w,m_w;x)\nonumber\\
&=&\sum_{m_w\in \mathbb Z}\oint \frac{dw}{2\pi i w}w^{-s}z^{-m_w}Z_M
Z_{v_+} Z_{v_-}\label{gxyz} \eeqn where \beqn
Z_{v_+}=\left(x^{(1-r'')} w^{-1}\right)^{|m_w|/2}P.E.\left(\frac{wx^{r''}- w^{-1} x^{(2-r'')}}{1-x^2} x^{|m_w|}\right)\nonumber\\
Z_{v_-}=\left(x^{(1-r'')} w\right)^{|m_w|/2}P.E.\left(\frac{w^{-1}x^{r''}-w x^{(2-r'')}}{1-x^2} x^{|m_w|}\right)\nonumber\\
Z_{M}=x^{1-2r}P.E.\left(\frac{x^{2r}-x^{2-2r}}{1-x^2}\right).\label{qqm}
\eeqn

where the R-charge of the $v_\pm$ fields $r''=N_f(1-r)-N_c+1=1-r$. Eq. (\ref{gxyz}) is the generalized
index of $U(1)$ theory with matter $M, v_{\pm}$ where $v_{\pm}$ has charge $\pm 1$ under $U(1)$.  And the
original $U(1)$ gauge symmetry becomes $U(1)$ topological symmetry of this new theory.
If we directly gauge XYZ model, the generalized index  has $w^s$ instead of $w^{-s}$ in (\ref{gxyz}),
which is again due to $S^2=C$.  \\
\\

\section{ Aharony duality for $SU(N_c)$ gauge theory with $N_f$ fundamental flavors}

\begin{longtable}{|c|cccc|}
\hline
Fields & $U(1)_{R}$ & $U(1)_A$ & $SU(N_f)$ & $SU(N_f)$ \\
\hline
$Q$ & $r$ & 1 & $\mathbf N_f$ & $\mathbf 1$ \\
$\tilde Q$ & $r$ & 1 & $\mathbf 1$ &    $\mathbf {\bar N_f}$    \\
$M$ & $2r$ & $2$ & $\mathbf N_f$ &   $\mathbf {\bar N_f}$   \\
$Y$ & $2N_f(1-r)-2N_c+2$ & $-2Nf$ & $\mathbf 1$ &    $\mathbf 1$    \\
\hline
$q$ & $1-r$ & $-1$ & $\mathbf {\bar N_f}$ &     $\mathbf 1$\\
$\tilde q$ & $1-r$ & $-1$ & $\mathbf 1$ &   $\mathbf N_f$   \\
$v_\pm$ & $N_f(1-r)-N_c+1$ & $-Nf$ & $\mathbf 1$ &   $\mathbf 1$    \\
$V_\pm$ & $N_f(r-1)+N_c+1$ & $N_f$ & $\mathbf 1$ &   $\mathbf 1$    \\
$u_\pm$ & $N_f(r-1)+N_c$ & $N_f$ & $\mathbf 1$ &    $\mathbf 1$ \\
\hline
\caption{The global symmetry charges of the chiral fields.}
\end{longtable}
Let's consider the Aharony duality for $SU(N_c)$ gauge theory with $N_f$ flavors.
Following the procedure of the previous section, we propose the following;
\begin{itemize}
\item
Electric theory: $SU(N_c)$ gauge theory(without Chern-Simons term),
$N_f$ pairs of fundamental/ anti-fundamental chiral superfields
$Q^a$, $\tilde{Q}_b$(where $a$, $b$ denote flavor indices).
\item
Magnetic theory: $U(1) \times U(N_f - N_c)$  gauge theory with the BF coupling
\begin{equation}
A_{U(1)}\wedge d {\rm Tr} A_{U(N_f-N_c)}  \label{BFU}
\end{equation},
with $N_f$ pairs of fundamental/anti-fundamental chiral
superfields $q_a$, $\tilde{q}^a$ of $U(N_f-N_c)$, $N_f \times N_f$ singlet
superfields $(M_{j})^{a}_{b}$, $j=0,\ldots,n-1$. We have $v_{\pm}, V_{\pm}$ charged under
$U(1)$ with charge $\pm 1$. The superpotential is given by
\begin{equation}
W=v_{+}V_{-}+v_{-}V_{+}+Mq\tilde{q}  \label{supergeneral}
\end{equation}
where $u_{\pm}$ is the monopole operator of $U(1)$.

\end{itemize}
 Note that the gauged fields $v_{\pm}$ do not have the usual $U(1)_R$ charge compared with
the elementary fields $Q, q$ since they have the nonperturbative origin. This will lead to interesting dynamics
such as the nonperturbatuve truncation of the chiral ring.

Let's compare the chiral ring elements of the both sides. For $SU(N_c)$
gauge theory with no superpontial, there are mesons $M^a_b=Q^a\tilde
Q_{\tilde b}$, a monopole operator $Y$ which parametrizes the Coulomb
branch, and baryons of the form \beqn B_{a_1\cdots
a_{N_f-N_c}}=\epsilon_{a_1\cdots a_{N_f-N_c}b_{1}\cdots
b_{N_c}}\epsilon^{i_{1}\cdots i_{N_c}}Q_{i_1}^{b_1}\cdots
Q_{i_{N_c}}^{b_{N_c}}. \eeqn
And, similary, there are $\tilde B\sim
\tilde Q^{N_c}$ in the chiral ring. Where $a,b$ and $i$ are the
flavor and gauge indices $a,b=1,\cdots,N_f$ and $i=1,\cdots,N_c$.
In the magnetic theory $Y$ is mapped to $v_{+}v_{-}$, mesons are mapped to
singlet fields $M$. Baryon fields $B, \tilde{B}$  are mapped to the monopole operators
$\tilde b\sim u_+\tilde
q^{a_1}\cdots \tilde q^{a_{\tilde N_c}}$ and $b\sim u_-q_{a_1}\cdots
q_{a_{\tilde N_c}}$ where the flavor and gauge indices are totally
anti-symmetric. The structure is a baryon operator of $SU(N_f-N_c)$ coupled
to the basic monopole operator $u_{\pm}$. This structure is required due to
the BF coupling. We can view the monopole operators as states on $S^2 \times R$ by operator-state correspondence of
 conformal field theories. When we turn on unit monopole of $A_{U(1)}$, due to
the BF coupling  (\ref{BFU}), $N_f-N_c$ matters should couple.\footnote{In our convention, each $q_{a}$ has the charge
 $\frac{1}{\tilde{N_c}}$ with $\tilde{N_c}= N_f-N_c$ under the overall $U(1)$ of $U(\tilde{N_c})$.} Due to the residual
gauge invariance of $SU(N_f-N_c)$ the allowed operator should have the
form $u_+\tilde q^{a_1}\cdots \tilde q^{a_{\tilde N_c}}$ and $
u_-q_{a_1}\cdots q_{a_{\tilde N_c}}$. Note that number of the baryons of both sides are the same $_{N_f}C_{N_c}=_{N_f}C_{N_f-N_c}=\frac{N_f!}{(N_f-N_c)!N_c!}$. $u_+u_-$ is  Q-exact since it has no BPS charge to protect.
 It is truncated nonperturbatively.\footnote{This is the special for $U(1)$ theory. For $U(N_c>1)$ theory,
 $u_{+}\sim (1,0,\cdots), u_{-}\sim (-1,0,\cdots)$ where we denote the monopole charge for Cartans $U(1)^{N_c}$ of $U(N_c)$.
 $u_{+}u_-$ is a BPS state.}

It's worthwhile to consider the special cases.  When $N_f=N_c\neq 1 $ the dual is
simply given by  $U(1)$ gauge theory with singlet $M$, unit charged
matter $v_{\pm}$ with the  superpotential
\begin{equation}
W=v_{+}v_{-}{\rm det} M  \label{superspecial}
\end{equation}
Note that the superpotential  is
inherited from the Aharony dual of $U(N_c)$ theory with $N_f=N_c$.
The electric theory has two baryon operators $B, \tilde{B}$. The
corresponding operator in the magnetic theory is given by $u_{+},
u_{-}$. For $N_f=N_c\neq 1$, the theory can also be described by mesons $M$, baryon fields $B, \tilde{B}$ and
monopole field $Y$ with the superpotential \cite{Seiberg3d}
\begin{equation}
W=-Y({\rm det} M-B\tilde{B})
\end{equation}
One can check that both theories have the same chiral ring structure and the same superconformal index.
Note that for the $U(1)$ theory with the superpotential (\ref{superspecial}), the  corresponding chiral ring relation
${\rm det} M-u_+u_-=0 $ should be generated by the $U(1)$ gauge dynamics.\footnote{This was pointed out by the authors of
\cite{Seibergsu} after the first version of the draft was distributed. We thank them for making this point clear.}

When $N_f=N_c=1$ the dual is $U(1)$ gauge theory with singlet $M$, charged matter $v_{\pm}$
with the superpotenial
\begin{equation}
W=v_{+}v_{-}{\rm det} M
\end{equation}
This was already discussed in the previous section.

Also interesting case is $N_c=1$ with arbitrary $N_f>1$. The
electric theory is free theory with $Q, \tilde{Q}$. The magnetic
theory is given by $U(1) \times U(N_f-1)$ belonging to the generic
case. The interesting thing is that we have $2N_f$ free fields $Q,
\tilde{Q}$. These are matched by $u_+\tilde q^{a_1}\cdots \tilde
q^{a_{\tilde N_f-1}}$ and $ u_-q_{a_1}\cdots q_{a_{\tilde N_f-1}}$.
Apparently the chiral ring element $v_{+}v_{-}$ exists in the
magnetic side, which has no counterpart in the electric side.
The truncation of this element occurs nonperturbatively.
This is similar to what happens to $N=2 \,\, U(1)$ theory with $N_f>1$ flavors.
The monopole operator $v'_+, v'_-$ in this theory has the same quantum number as the above
$v_{+}, v_{-}$. Since the product of $v'_+v'_-$ has no BPS charge to protect, it is
truncated nonperturbatively. We think the similar thing happens in the case at hand as well.
Indeed one can see this operator is canceled by a suitable fermionic operator
in the index computation.

Of course when $N_f=1$, $v_+,v_-$ has the quantum number of elementary fields.
In this case due to  the usual superpotential $W=v_+v_-M$,  $v_+v_-$ is Q-exact.

\subsection{Index of $SU(N_c)$ theory obtained from ungauging $U(N_c)$}

Now let's consider the index of the electric $U(N_c)$ theory with
$N_f$ flavors.
\beqn I(w,m_w;x)=\sum_{m_i\in \mathbb
Z}\frac{1}{Sym}\oint \prod_{i=1}^{N_c}\frac{dz_i}{2\pi
iz_i}w^{m_1+\ldots+m_{N_c}}(z_1\ldots
z_{N_c})^{m_w}Z_{gauge}Z_{Q}^{N_f}Z_{\tilde Q}^{N_f}.\label{eindex2}
\eeqn where $w$ is the fugacity for $U(1)_T$ and $z_i$ denotes the
holonomy variable of the Cartans of $U(N)$. $z=z_1z_2 \cdots
z_{N_c}$ is the holonomy variables of the overall $U(1)$, $Z_{Q}$
and $Z_{\tilde Q}$ is given by
\beqn
Z_{Q}=\prod_{i=1}^{N_c}\left(x^{1-r}z_i^{-1}\right)^{|m_i|/2}P.E.
\left(\frac{z_i x^r-z_i^{-1}x^{2-r}}{1-x^2}x^{|m_i|}\right)\nonumber\\
Z_{\tilde
Q}=\prod_{i=1}^{N_c}\left(x^{1-r}z_i\right)^{|m_i|/2}P.E.\left(\frac{z_i^{-1}x^r-z_ix^{2-r}}{1-x^2}x^{|m_i|}\right).
\label{qq} \eeqn And $Z_{gauge}$ is \beqn
Z_{gauge}=x^{-\sum_{i<j}^{N_c}|m_i-m_j|}\prod_{i<j}^{N_c}(1-z_i
z_j^{-1}x^{|m_i-m_j|})(1-z_i^{-1} z_jx^{|m_j-m_i|}) \label{gauge}
\eeqn

Equivalently this can be viewed as gauging overall $U(1)$ global
symmetry of $SU(N_c)$ with additional BF term $A_{new} \wedge
dA_{U(1)}$ where $U(1)$ symmetry acts as $Q_i\rightarrow e^{i\theta}
Q_i$. Thus quark has charge 1 under this $U(1)$ symmetry. Thus the
same index can be written as
\begin{equation}
 I(w,m_w;x)=\sum_{m_\in \mathbb
Z}\oint \frac{dz}{2\pi iz} w^s z^m I_{SU(N_c)}(z,s)
\end{equation}
where $I_{SU(N_c)}(z,s)$ is the generalized index of $SU(N_c)$
theory. Comparing it with eq. (\ref{eindex2}) we find that
\begin{equation}
z=z_1z_2 \cdots z_{N_c}, \,\,\,\, s=m_1+m_2+\cdots m_{N_c}.
\label{surel}
\end{equation}

Let's just
denote that $\frac{1}{sym}Z_{gauge}Z_{Q}^{N_f}Z_{\tilde
Q}^{N_f}=Z_{\Phi}(m_1,z_1,m_2,z_2,\cdots,m_{N_c},z_{N_c};x)$ for
simplicity. Concentrating on an arbitrary $z_i$ out of $z_1,\cdots,
z_{N_c}$, say  $z_i=z_{N_c}$, we can expand
$Z_{\Phi}=\sum_{n\in\mathbb Z}z_{N_c}^n \tilde
Z_{\Phi}(\cdots,m_{N_c},n)$. Then by rewriting  \eqref{eindex2}
using this expansion of $z_{N_c}$, and ungauing $U(1)$ of $U(N_c)$, it
becomes \beqn
\tilde{I}(z,s)&=&\sum_{m_w\in \mathbb Z}z^{-m_w}\oint \frac{dw}{2\pi i w^{s+1}}I(w,m_w;x)\nonumber\\
&=&\sum_{m_w,m_i\in \mathbb Z}\oint \frac{dw}{2\pi i
w}\left(\prod_{i=1}^{N_c}\frac{dz_i}{2\pi
iz_i}\right)w^{m_1+\ldots+m_{N_c}-s}\left(\frac{z_1\ldots
z_{N_c}}{z}\right)^{m_w}\sum_{n\in\mathbb Z}z_{N_c}^n \tilde
Z_{\Phi}. \nonumber \eeqn By integrating $w$ and $z_{N_c}$, we obtain
 \beqn
\tilde{I}(z,s)=\sum_{m_i\in\mathbb Z}\oint
\left(\prod_{i=1}^{N_c-1}\frac{dz_i}{2\pi iz_i}\right)
\sum_{m_w\in\mathbb Z}\left(\frac{z}{z_1\ldots z_{N_c-1}}
\right)^{m_w} \tilde Z_{\Phi}(\cdots,s-m_1-\cdots-m_{N_c-1},m_w). \nonumber
\eeqn \indent This means that the $z_{N_c}$ is changed by the
$\left(\frac{z}{z_1\ldots z_{N_c-1}}\right)$ and the $m_{N_c}$ is
changed by the $s-m_1-\cdots-m_{N_c-1}$ from the origianl $U(N_c)$
gauge theory, which we already find at eq. (\ref{surel}). Thus we
recover the generalized index of $SU(N_c)$ gauge theory with the
global $U(1)$ with fugacity $z$ and charge $s$, as expected.
By setting $z=1$ and $s=0$, we obtain the ungauged index $\tilde
I(1,0)$, which is equal to the index of usual $SU(N_c)$ theory with $N_f$ flavors.

\subsection{Gauging $U(1)_T$ of magnetic $U(N_f-N_c)$ theory}
Index of magnetic thoery with  $U(N_f-N_c=\tilde N_c)$ gauge theory
is \beqn I=\sum_{m_i\in \mathbb Z}\frac{1}{Sym}\oint
\prod_{i=1}^{\tilde N_c}\frac{dz_i}{2\pi
iz_i}w^{m_1+\ldots+m_{N_c}}(z_1\ldots
z_{N_c})^{m_w}Z_{gauge}Z_{q}^{N_f}Z_{\tilde
q}^{N_f}Z_{M}^{N_f^2}Z_{v_+}Z_{v_-}. \eeqn\label{mindex2} Where
\beqn
Z_{q}=\prod_{i=1}^{\tilde N_c}\left(x^{(1-r')}z_i^{-1}\right)^{|m_i|/2}P.E.\left(\frac{z_i x^{r'}-z_i^{-1} x^{(2-r')}}{1-x^2} x^{|m_i|}\right)\nonumber\\
Z_{\tilde{q}}=\prod_{i=1}^{\tilde N_c}\left(x^{(1-r')}z_i\right)^{|m_i|/2}P.E.\left(\frac{z_i^{-1}  x^{r'}-z_i x^{(2-r')}}{1-x^2} x^{|m_i|}\right)\nonumber\\
Z_{M}=x^{1-2r}P.E.\left(\frac{x^{2r}-x^{2-2r}}{1-x^2}\right)\nonumber\\
Z_{v_+}=\left(x^{(1-r'')}a w^{-1}\right)^{|m_w|/2}P.E.\left(\frac{wx^{r''}- w^{-1} x^{(2-r'')}}{1-x^2} x^{|m_w|}\right)\nonumber\\
Z_{v_-}=\left(x^{(1-r'')}
w\right)^{|m_w|/2}P.E.\left(\frac{w^{-1}x^{r''}-w
x^{(2-r'')}}{1-x^2} x^{|m_w|}\right).\label{qqm} \eeqn
The R-charge of chiral field $q$ is $r'=1-r$ and R-charge of $v_{\pm}$ is $r''=N_f(1-r)-N_c+1$.\\

\indent Let's ungauge $U(1)_T$ of the magnetic theory \beqn
\tilde{I}(z,s)&=&\sum_{m_w\in \mathbb Z}z^{-m_w}\oint \frac{dw}{2\pi i w^{s+1}}I(w,m_w;x)\nonumber\\
&=&\sum_{m_w\in\mathbb Z}\oint \frac{dw}{2\pi i
w}w^{m_1+\ldots+m_{N_c}-s}\left(\frac{z_1\ldots z_{\tilde
N_c}}{z}\right)^{m_w}Z_{v_+}Z_{v_-}\times(\cdots) \eeqn Where the
$\times(\cdots)$ term is independent of $w$ and $m_w$. Hence we can find $v_{\pm}$ are gauged in the index expression.
Terms denoted above are just the form of the $U(1)$ gauge theory
with fugacity $w$, charge $m_w$. Also from the functional form of  $w, z$ one can see that there is a BF coupling
between gauged $U(1)_T$  and overall $U(1)$ factor of
$U(N_f-N_c)$.

\subsection{Results of Indices}

We can check the  dualities by using superconformal indices. We
expanded the indices about $x$ upto some orders and compared the
dual pairs for $0<N_c ,\tilde N_c <3$. We turned off the ungauged
global $U(1)$ symmetry by setting fugacity variable $z=1$ and charge
$s=0$.

\begin{longtable}{|c|c|c|p{7.5cm}|}
\hline
  &  Electric  &  Magnetic  &  \\
$(N_c,N_f)$  &  $SU(N_c)$  &  $U(1)\times U(N_f - N_c)$  &  Index ($r$ is the IR $R$-charge of $Q$.)\\
\hline
$(1,1)$       &   free   &  $U(1)$    &  $1-4 x^2-5 x^4+x^{4-2 r}-2 x^{2-r}+2 x^r+3 x^{2 r}+4 x^{3 r}+5 x^{4 r}-4 x^{2+r}-4 x^{2+2 r}+\cdots$  \\
\hline
$(1,2)$       &   free   &  $U(1)\times U(1)$      &  $1-16 x^2+6 x^{4-2 r}-4 x^{2-r}+4 x^r+10 x^{2 r}+20 x^{3 r}-36 x^{2+r}+\cdots$   \\
\hline
$(1,3)$       &   free   &  $U(1)\times U(2)$      &  $1-36 x^2+15 x^{4-2 r}-6 x^{2-r}+6 x^r+21 x^{2 r}+56 x^{3 r}-120 x^{2+r}+\cdots$   \\
\hline
$(1,4)$       &   free   &  $U(1)\times U(3)$      &   $1-64 x^2+28 x^{4-2 r}-8 x^{2-r}+8 x^r+36 x^{2 r}+120 x^{3 r}-280 x^{2+r}+\cdots$  \\
\hline
$(2,2)$       &   $SU(2)$    &  $U(1)$         &  $1-16 x^2+88 x^4+x^{4-8 r}+x^{2-4 r}+26 x^{4-2 r}+6 x^{2 r}+20 x^{4 r}+50 x^{6 r}+105 x^{8 r}-64 x^{2+2 r}-160 x^{2+4 r}+\cdots$   \\
\hline
$(2,3)$       &   $SU(2)$   &  $U(1)\times U(1)$         &  $1-36 x^2+558 x^4+x^{8-12 r}+x^{4-6 r}+21 x^{4-2 r}+15 x^{2 r}+105 x^{4 r}+490 x^{6 r}-384 x^{2+2 r}+\cdots$   \\
\hline
$(2,4)$       &   $SU(2)$   &  $U(1)\times U(2)$         &  $1-64 x^2+1888 x^4+x^{12-16 r}+x^{6-8 r}+36 x^{4-2 r}+28 x^{2 r}+336 x^{4 r}-1280 x^{2+2 r}+\cdots$   \\
\hline
$(2,5)$       &   $SU(2)$   &  $U(1)\times U(3)$       &  $1-100 x^2+4750 x^4+x^{16-20 r}+x^{8-10 r}+55 x^{4-2 r}+45 x^{2 r}+825 x^{4 r}-3200 x^{2+2 r}+\cdots$  \\
\hline
$(3,3)$       &   $SU(3)$   &   $U(1)$      &  $1+82 x^2+x^{2-6 r}+9 x^{2-4 r}+36 x^{2-2 r}+9 x^{2 r}+2 x^{3 r}+45 x^{4 r}+18 x^{5 r}+167 x^{6 r}+90 x^{7 r}+513 x^{8 r}+332 x^{9 r}-18 x^{2+r}+81
x^{2+2 r}-162 x^{2+3 r}+\cdots$   \\
\hline
$(3,4)$       &   $SU(3)$   &  $U(1)\times U(1)$         &   $1-32 x^2+x^{4-8 r}+16 x^{4-6 r}+100 x^{4-4 r}+16 x^{2 r}+8 x^{3 r}+136 x^{4 r}+120 x^{5 r}+836 x^{6 r}-48 x^{2+r}-480 x^{2+2 r}+\cdots$  \\
\hline
$(3,5)$       &   $SU(3)$   &  $U(1)\times U(2)$         &   $1-50 x^2+x^{6-10 r}+25 x^{6-8 r}+25 x^{2 r}+20 x^{3 r}+325 x^{4 r}+450 x^{5 r}-100 x^{2+r}+\cdots$  \\
\hline
$(3,6)$       &   $SU(3)$   &  $U(1)\times U(3)$         &   $1-72 x^2+x^{8-12 r}+36 x^{8-10 r}+36 x^{4-2 r}+36 x^{2 r}+40 x^{3 r}+666 x^{4 r}-180 x^{2+r}+\cdots$  \\
\hline
\caption{The results of the superconformal index computation.}\label{index}
\end{longtable}

It's worthwhile to examine the chiral ring structure of the theories
and to work out the gauge invariant operators of the first few
lower orders of the indices. For the case of the electric  theory,
$SU(N_c)$ gauge theory with no superpotential, there are mesons
$M^a_b=Q^a\tilde Q_{\tilde b}$, a monopole operator $Y$, and baryons
of the form $B \sim Q^{N_c}$ , $\tilde B\sim \tilde Q^{N_c}$ with  $a,b$  the flavor  indices
$a,b=1,\cdots,N_f$. The lowest components of
meson superfields give $N_f^2x^{2r}$ to the index. From the bosonic components
$B$ and $\tilde B$,  we pick $N_c$ of chiral fields among
$N_f$ so that it gives $_{N_f}C_{N_c}x^{N_c r}$ from each $B$ and $\tilde
B$.\footnote{$_{N_f}C_{N_c}={N_f\choose N_c}=\frac{N_f!}{N_c!(N_f-N_c)!}$} For example, for $N_c=2$, indices have the contribution
 \beqn
I=\cdots+\left(N_f^2+N_f(N_f-1)\right)x^{2r}+\cdots \eeqn and, for
$N_c=3$, indices have
\beqn
I=\cdots+N_f^2x^{2r}+\frac{1}{3}N_f(N_f-1)(N_f-2)x^{3r}+\cdots.
\eeqn
 There is a monopole operator $Y$ which corresponds to $v_+ v_-$ of the dual theory. Sinces its R charge is
 just the twice of $v_{\pm}$, it gives $x^{2N_f(1-r)+2-2N_c}$ to the indices. For instance, when $N_c=2$, $N_f=3$
 there is a term $x^{4-6r}$ for $Y$, $x^{8-12r}$ for $Y^2$\\
 \indent As pointed out in \cite{Seiberg3d} for the case $N_f=N_c$, the theory  is described by in terms of $M, Y, B, \tilde{B}$
 with the superpotential
\beqn
W=-Y(\mathrm{det}M-B\tilde B)
\eeqn
which gives constraint that $\mathrm{det}M-B\tilde B=0$. One state corresponding to them becomes Q-exact so it doesn't appear
in the index result. We can check it in the index expansion. From $\mathrm{det}M$ we
have $_{N_f^2}\mathrm{H}_{N_f}x^{2 N_f r}$\footnote{$_{n}H_{m}=_{n+m-1}C_{m}=\frac{(n+m-1)!}{(n-1)!m!}$},
and from $B^2$, $\tilde B^2$,$B\tilde{B}$,  $MB$ and $M\tilde B$ we have $(3+2N_f^2)x^{2 N_f r}$.
But actually the coefficient of the term is smaller by 1 than the naive counting.

There are terms like $x^2$. This is due to the mixed contribution of bosonic and fermionic operators. For example consider $N_c=1$ case. Because these are free chiral theories so
just $Q$ and $\tilde Q$ are chiral ring elements.  $Q$ and  $\tilde Q$ contribute $2N_f x^r$ to the index.  Fermion components of chiral fields give $-2N_f x^{2-r}$, and to the next order,
two chiral fields give $_{2N_f}H_{2}x^{2r}$. When fermions coupled to $Q$, $\tilde Q$ they gives $(2N_f)^{2}$ copies of $x^r\times x^{2-r}$ so there are $-(2N_f)^2 x^2$.

 As alluded before,  the chiral ring elements of magentic theory  corresponding to $B$ and
$\tilde B$ are
$\tilde b\sim u_+\tilde q^{a_1}\cdots \tilde q^{a_{\tilde N_c}}$ and
$b\sim u_-q_{a_1}\cdots q_{a_{\tilde N_c}}$ where the flavor and gauge indices are totally anti-symmetric.
 Since $u_{\pm}$ and $q$,
$\tilde q$ have R-charge $r_u=N_f(1-r)-N_c$ and $1-r$ respectively, $b$ and
$\tilde b$ have R-charge $N_f(1-r)-N_c+\tilde N_c(1-r)=N_c r$. Since we
pick $\tilde N_c=N_f-N_c$ of $q$ and $\tilde q$ out of $N_f$, $b$ and $\tilde b$
together give $2\times{N_f\choose N_f-N_c}x^{N_c r}$ to the index. \\
\indent For the case of $N_c=1$, on the other hand, there is no
monopole operator in electric theories because they are just free
chiral field theories. But there are still $v_{\pm}$ exist in the
dual theories which possibly can make the chiral ring element $v_+v_-$.
We propose that this chiral ring element is truncated nonperturbatively.
 $v_-v_+$ gives $x^{N_f(2-2r)}$ but it is cancled by the contribution of fermionic partner of   $\mathrm{det}M$ which gives $-x^{N_f(2-2r)}$.\\
\indent When
$N_f=N_c=1$, the magnetic theory is $U(1)$ theory with 3 fields $v_{\pm},M$ with superpotential $W=v_+v_-M$.   $v_{\pm}$ give $2x^r$ to the index. Due to
the superpotential $W=v_+v_-M$, $v_+v_{-}$ are
Q-exact. So the singlet $M$ gives $x^{2r}$
to the index and monopoles from charge sector of $2$ and $-2$,
$u_+^2$ and $u_-^2$ respectively, give $2x^{2r}$ to the index.
In sum, we have $3x^{2r}$. One can go to higher orders if one wishes.

\section{Chern-Simons theory of $SU(N_c)_k$ and it's dual}
In the standard fashion, the duality of  Chern-Simons theory of $SU(N_c)$ gauge theory can be
obtained from the Aharony duality of $SU(N_c)$ gauge theory. By
giving some of the flavors axial mass, one can generate Chern-Simons
term. We start with $SU(N_c)$ theory with $N_f+k$ flavors. Matters
have axial charge $+1$ and integrating of $k$ matters gives  CS
level $k>0$. The dual theory starts with $U(1)\times U(N_f+k-N_c)$
gauge theory with BF term $A_{U(1)}\wedge d {\rm Tr} A$, where ${\rm Tr} A$
denotes the overall $U(1)$ of $U(N_f+k-N_c)$. Integrating $k$ of
$q$, $\tilde q$ gives CS $-k$ of $U(N_f+k-N_c)$ , since $q$, $\tilde q$ have axial charge $-1$.
Integrating out $v_{\pm}$ gives CS level $-1$ for $U(1)$ gauge
field. Thus the dual theory is given by $U(1)_{-1} \times
U(N_f+k-N_c)_{-k}$ with $N_f$ flavors only charged under
$U(N_f+k-N_c)$ and the BF term $A_{U(1)}\wedge d {\rm Tr} A$. Subscript of
the gauge group denotes the Chern-Simons level. Note that the above
theory is charge-conjugation invariant. Now no field is charged
under $U(1)_{-1}$ so it can be integrated out. Thus we obtain the
following duality;
\begin{itemize}
\item
Electric theory: $SU(N_c)_k$ gauge theory, $N_f$ pairs of
fundamental/ anti-fundamental chiral superfields $Q^a$,
$\tilde{Q}_b$(where $a$, $b$ denote flavor indices) with
Chern-Simons level $k$.
\item
Magnetic theory: $U(N_f+k- N_c)$  gauge theory, $N_f$ pairs of
fundamental/anti-fundamental chiral superfields $q_a$, $\tilde{q}^a$
of $U(N_f+k-N_c)$, $N_f \times N_f$ singlet superfields
$(M_{j})^{a}_{b}$, $j=0,\ldots,n-1$. The superpotential is given by
\begin{equation}
W=Mq\tilde{q}.
\end{equation}
with the Chern-Simons term
\begin{equation}
\tilde{A} \wedge d \tilde{A}-k (A_{N_{\tilde{c}}} \wedge d
A_{N_{\tilde{c}}} -\frac{2i}{3} A_{N_{\tilde{c}}}\wedge
A_{N_{\tilde{c}}} \wedge A_{N_{\tilde{c}}})
\end{equation}
\end{itemize}
where $A_{N_{\tilde{c}}}$ is the $U(N_{\tilde{c}})=U(N_f+k-N_c)$ gauge field and $\tilde{A}={\rm Tr} A_{N_{\tilde{c}}}$ is the overall $U(1)$ gauge field.

Again we can discuss the chiral ring elements. In the elctric side
we have mesons $M$, and baryons $B, \tilde{B}$. In the magnetic side,
mesons are trivially matched. The baryon operators are mapped to the
monopole operators of the form $ u_+\tilde q^{a_1}\cdots \tilde
q^{a_{\tilde N_f-N_c}}$ and $ u_-q_{a_1}\cdots q_{a_{\tilde
N_f-N_c}}$. If we turn on the unit flux of the overall $U(1)$ of
$U(N_f+k-N_c)$, Gauss constraint dictates that we should turn on
$N_f-N_c$ matters.  Using the residual gauge invariance, one can
choose the color index running from $1$ to $N_f-N_c$.

\indent The index of Chern-Simons $SU(N_c)$ theory is \beqn
I=\sum_{m_i\in \mathbb Z}\frac{1}{Sym}\oint
\prod_{i=1}^{N_c}\frac{dz_i}{2\pi iz_i}(-z_i)^{k m_i}
Z_{gauge}Z_{Q}^{N_f}Z_{\tilde Q}^{N_f}, \eeqn where $Z_{Q}$,
$Z_{\tilde Q}$, $Z_{gauge}$ are given by eq. (\ref{qq}),
(\ref{gauge}).
Here $z_{N_c}=(z_1\cdots z_{N_c-1})^{-1}$ and $m_{N_c}=-(m_1+\cdots+m_{N_c-1})$.\\
\indent The index of dual Chern-Simons $U(N_f+k-N_c)$ is \beqn
I=\sum_{m_i\in \mathbb Z}\frac{1}{Sym}\oint \prod_{i=1}^{\tilde
N_c}(-1)^{-k m_i+m_i}z_i^{-km_i+m_1+\cdots+m_{\tilde
N_c}}\frac{dz_i}{2\pi iz_i}Z_{gauge}Z_{q}^{N_f}Z_{\tilde
q}^{N_f}Z_{M}^{N_f^2}, \label{dualindex}\eeqn where the same
expressions are adopted from (\ref{qqm}).

Here the $ N_{\tilde{c}}$ is the rank of the gauge group $
N_{\tilde{c}}=N_f+k-N_c$. Note that there's additional sign factor $(-1)^{-k
m_i+m_i}$ at eq. (\ref{dualindex}). Such sign factor $(-1)^{km}$ is
observed for every $U(1)$ factor with CS level $k$ where $m$ is the
magnetic flux associated with $U(1)$ gauge group. The origin of such sign factor is described
in \cite{DGG2}.
 \\
\indent We compute the index of the above CS theory  for some cases and find the perfect matchings.
\begin{longtable}{|c|c|c|p{7.5cm}|}
\hline
  &  Electric  &  Magnetic  &  \\
$(N_c,N_f,k)$  &  $SU(N_c)$  &  $U(N_f +k - N_c)$  &  Index ($r$ is the IR $R$-charge of $Q$.)\\
\hline
$(2,2,1)$       &  $SU(2)$   &  $U(1)$    & $1-16 x^2+88 x^4-x^{4-4 r}+20 x^{4-2 r}+6 x^{2 r}+20 x^{4 r}-64 x^{2+2 r}+\cdots$  \\
\hline
$(2,1,2)$       &   $SU(2)$   &  $U(1)$    & $1-4 x^2-5 x^4+4 x^6+3 x^{4-2 r}+4 x^{6-2 r}+x^{2 r}+x^{4 r}+x^{6 r}-4 x^{4+2 r}+\cdots$  \\
\hline
$(2,3,1)$       &   $SU(2)$   &  $U(2)$    &  $1-36 x^2+558 x^4-x^{6-6 r}+21 x^{4-2 r}+15 x^{2 r}+105 x^{4 r}+490 x^{6 r}-384 x^{2+2 r}+\cdots$ \\
\hline
$(2,2,2)$       &   $SU(2)$   &  $U(2)$    &  $1-16 x^2+88 x^4+10 x^{4-2 r}+6 x^{2 r}+20 x^{4 r}+50 x^{6 r}+105 x^{8 r}-64 x^{2+2 r}-160 x^{2+4 r}+\cdots$ \\
\hline
$(2,4,1)$       &   $SU(2)$   &  $U(3)$    &  $1-64 x^2+36 x^{4-2 r}+28 x^{2 r}+336 x^{4 r}+\cdots$ \\
\hline
$(2,3,2)$       &   $SU(2)$   &  $U(3)$    &  $1-36 x^2+15 x^{2 r}+105 x^{4 r}+\cdots$ \\
\hline
$(3,3,1)$       &   $SU(3)$   &  $U(1)$    &  $1-18 x^2+9 x^{2 r}+2 x^{3 r}+45 x^{4 r}+18 x^{5 r}+167 x^{6 r}+90 x^{7 r}+513 x^{8 r}+332 x^{9 r}-18 x^{2+r}-144 x^{2+2 r}-162 x^{2+3 r}+\cdots$ \\
\hline
$(3,2,2)$       &   $SU(3)$   &  $U(1)$    &  $1-8 x^2+28 x^4-x^{4-4 r}+8 x^{4-2 r}+16 x^{4-r}+4 x^{2 r}+10 x^{4 r}-4 x^{2+r}-24 x^{2+2 r}+\cdots$ \\
\hline
$(3,4,1)$       &   $SU(3)$   &  $U(2)$    &  $1-32 x^2+16 x^{2 r}+8 x^{3 r}+136 x^{4 r}+120 x^{5 r}+836 x^{6 r}-48 x^{2+r}-480 x^{2+2 r}+\cdots$ \\
\hline
$(3,3,2)$       &   $SU(3)$   &  $U(2)$    &  $1-18 x^2+9 x^{2 r}+2 x^{3 r}+45 x^{4 r}+18 x^{5 r}+167 x^{6 r}+90 x^{7 r}+513 x^{8 r}+332 x^{9 r}-18 x^{2+r}-144 x^{2+2 r}-162 x^{2+3 r}+\cdots$ \\
\hline
$(3,2,3)$       &   $SU(3)$   &  $U(2)$    &  $1-8 x^2+4 x^{4-2 r}+4 x^{2 r}-4 x^{2+r}+\cdots$ \\
\hline
\caption{The results of the superconformal index comutation of Chern-Simons theory}\label{index}
\end{longtable}

We examine how gauge invariant BPS operators appear on
the index expression. On the electric theory side, since these are
Chern-Simons $SU(N)$ gauge theories, chiral ring elements
consist of mesons $M$ and  baryons  $B$, $\tilde B$.
Mesons $M=Q\tilde Q$ contribute $N_f^2 x^{2r}$  to the index  and
baryons $B\sim
Q^{N_c}$, $\tilde B\sim \tilde{Q}^{N_c}$ give $2\times{N_f\choose N_c}x^{N_c r}$ to the index.
But when $N_f<N_c$ there are no baryons. It is easily seen from the result we have here.
 For example, when $N_c=2$, we should look for the coefficient of $x^{2r}=x^{N_c r}$ term.
 If $N_f=2$ there is the baryon contribution $I=\cdots+(2^2+2{2\choose 2})x^{2r}+\cdots$ but
  if $N_f=1$ there is no such  contribution.\\
\indent Looking for magnetic theory,  mesons are trivially mapped to
the siglet operator $M$  the magnetic theory always have. As explained before, baryon operators are matched by the monopole
operators $u_+ \tilde q^{a_{\tilde N_f-N_c}}$ and $ u_-q_{a_1}\cdots
q_{a_{\tilde N_f-N_c}}$.  They give ${N_f\choose N_f-N_c}$ for each
$\pm 1$ monopole flux, which give the term $2{N_f\choose
N_f-N_c}x^{N_c r}$.

\section{$SU(N)$ theory with $N_f$ fundamental flavors and adjoint matter}

\begin{longtable}{|c|c|c|c|c|l|}
\hline
             & $SU(N_f)$   & $SU(N_f)$ & $U(1)_A$ & $U(1)_J$& $U(1)_{R}$      \\
\hline
$Q$          & ${\bf N_f}$ & ${\bf 1}$ & 1 & 0 & $r$                        \\
${\tilde Q}$ & ${\bf 1}$   & ${\bf \overline{N_f}}$& 1  & 0& $r$              \\
$X,\tilde X$        & ${\bf 1}$   & ${\bf 1}$ & 0 & 0 & $\frac{2}{n+1}$            \\
$M_j$        & ${\bf N_f}$ & ${\bf \overline{N_f}}$& 2   & 0& $2r+\frac{2j}{n+1}$  \\
$v_{j,\pm}$  & ${\bf 1}$   & ${\bf 1}$& $-N_f$  & $\pm 1$& $-N_fr+N_f-\frac{2}{n+1}(N_c-1)+\frac{2j}{n+1}$ \\
$q$          & ${\bf \overline{N_f}}$ & ${\bf 1}$& -1  & 0& $-r+\frac{2}{n+1}$   \\
${\tilde q}$ & ${\bf 1}$ & ${\bf N_f}$& -1  & 0& $-r+\frac{2}{n+1}$         \\
$\tilde{v}_{j,\pm}$  & ${\bf 1}$ & ${\bf 1}$& $N_f$  & $\pm 1$& $N_fr-N_f+\frac{2}{n+1}(N_c+1)+\frac{2j}{n+1}$ \\
$u_\pm$          & ${\bf 1}$ & ${\bf 1}$ & $N_f$ & 0 & $-nN_f(1-r)+\frac{2n}{n+1}(N_c-1)+\frac{2n}{n+1}$   \\
\hline
\caption{Quantum numbers of various fields}
\end{longtable}

One can apply the same procedure to obtain Aharony duality for $N=2 \,\,SU(N_c)$ gauge theory with $N_f$ fundamental and anti-fundamental flavors and with one adjoint matter. The proposed duality is as follows;
\begin{itemize}
\item
Electric theory: $SU(N_c)$ gauge theory(without Chern-Simons term),
$N_f$ pairs of fundamental/ anti-fundamental chiral superfields
$Q^a$, $\tilde{Q}_b$(where $a$, $b$ denote flavor indices), an
adjoint superfield $X$, and the superpotential $W_e=\Tr\, X^{n+1}$.
\item
Magnetic theory: $U(1) \times U(n N_f - N_c)$  gauge theory with BF coupling,
\begin{equation}
A_{U(1)}\wedge d {\rm Tr} A_{U(n N_f-N_c)}
\end{equation}
 $N_f$ pairs of fundamental/anti-fundamental chiral
superfields $q_a$, $\tilde{q}^a$ of $U(n N_f-N_c)$, $N_f \times N_f$ singlet
superfields $(M_{j})^{a}_{b}$, $j=0,\ldots,n-1$,  an adjoint superfield
$\tilde{X}$ of $U(n N_f-N_c)$ , $2n$
superfields $v_{0,\pm}$,\ldots,$v_{n-1,\pm}$ charged under $U(1)$,
$2n$
superfields $\tilde{v}_{0,\pm}$,\ldots,$\tilde{v}_{n-1,\pm}$ charged under $U(1)$
and a superpotential
\begin{equation}
W_m=\Tr\, \tilde{X}^{n+1}+\sum_{j=0}^{n-1} M_j
\tilde{q} Y\tilde{X}^{n-1-j} q
+\sum_{i=0}^{n-1}(v_{i,+}\tilde{v}_{n-1-i,-}+v_{i,-}\tilde{v}_{n-1-i,+}).
\end{equation}
\end{itemize}
The chiral superfields of the theory have charges under various symmetries as we specified at the table above.

 $v_{0,\pm}$ and $\tilde{v}_{0,\pm}$ are minimal bare monopoles
of electric theory and magnetic theory, respectively. Those
correspond to excitation of magnetic flux $(\pm 1,0,\ldots,0)$. For
the description of the monopole operators we had better use the
operator state correspondence to describe the operator as the
corresponding state on $R\times S^2$. When magnetic flux $(\pm
1,0,\ldots,0)$ is excited the gauge group $U(N_c)$ is broken to
$U(1)\times U(N_c-1)$.
Let \begin{equation}
X=\left( \begin{array}{cc}
               X_{11} & 0\\
               0      & X^{'}      \end{array}  \right)
\end{equation}
where $X^{'}$ is an adjoint field of $U(N_c-1)$ unbroken gauge
group. We denote the dressed monopole operator
$v_{i,\pm}\equiv \Tr(v_{0,\pm} X^i)$, $i=1,\ldots,n-1$ with the
trace taken over $U(1)$. For example $v_{1, \pm}=X_{11}\ket{\pm1,0,\ldots}$.
The details can be found in \cite{jparkadjoint}.

Let's consider the chiral ring elements and how they are mapped .
For the adjoint and mesons, we have the following correspondence
\beqn
\mathrm{Tr} X^i & \leftrightarrow & \mathrm{Tr} \tilde X^i\nonumber\\
Q^a X^{j} \tilde{Q}_b & \leftrightarrow &(M_{j})^{a}_{b}.
\eeqn
Where the $a$ and $b$ are the flavor indicies. We have $n$ independent monopole operators $Y_i$ with $i=0 \cdots n-1$, which are
mapped to
$ v_{i, +}v_{i,-}$.
 Baryons can be constructed not only from $Q$, $\tilde Q$ but also from some combinations of $X$ and $Q$ or $X$ and $\tilde Q$. For example $X^j_iQ^i$ or $X^j_iX^i_kQ_k$ can replace one or many $Q$ in the baryon operators, here $i, j, k$ are gauge indicies. Thus $B'\sim Q^{N_c-1}(XQ)$ could be a chiral ring element.

On the dual side,  baryon like operators $b$, $\tilde b$ which is a coupled state of the monopole operators $u_{\pm}$ of the gauged $U(1)$ and (anti)fundamental and adjoint matters fields.
 Where baryons could be $b\sim u_-q^{\tilde{N}_c}$ or $b'\sim u_- q^{\tilde{N_c}-1}(\tilde Xq)$ where $\tilde{N}_c=nN_f-N_c$.
 Anit-baryons $\tilde b$ could be defined  similarly,
 e.g., $\tilde b\sim u_+\tilde q^{\tilde{N_c}}$ or $\tilde b'\sim u_+ \tilde q^{\tilde{N_c}-1}(\tilde Xq)$.
 The detailed matching of the chiral ring is quite delicate. For the Aharony duality of $U(N_c)$ gauge theory with an adjoint
 $X$,
the chiral rings are constrained by characteristic equations of
adjoint $X$ and $\tilde{X}$. Classically, there are $N_c$ independent
operators $\Tr\,X^i$, $i=1,\ldots,N_c$ due to characteristic
equation of $X$ which is in $U(N_c)$ adjoint representation. With a
superpotential $W=\Tr\,X^{n+1}$ there are $a$ independent operators
$\Tr\,X^i$, $i=0,\ldots,a$ where $a= \textrm{min}(n-1, N_c)$. As explained in \cite{KapustinPark, jparkadjoint},
if the gauge group of electric side is smaller than
that of magnetic side, $N_c\leq N_c^{'}$, the number of (classical)
chiral ring generators of electric side is less than magnetic side.
The redundant chiral ring generators of magnetic side are
cancelled by some monopole operators.
On the other hand, if $N_c>N_c^{'}$, the electric theory seems to
have more chiral ring generators than magnetic theory. But some
non-trivial relation of monopole operators reduce the number of
state so the chiral ring is again the same.

 We expect similar mechanism works here.
Especially interesting case is baryon operators. For the electric case, such
baryon operators exist for $N_f\geq N_c$. However, for the magnetic side,
the condition is $N_f \geq \tilde{N}_c=nN_f-N_c$. Unless $n=2$ and $N_f=N_c$
two conditions are incompatible. Thus we expect the nonperturbative truncation of the chiral ring
occurs. Here we examine simple cases of $n=2$ and $N_f=N_c$ in the below and find the perfect matching.
We will explore more general cases in the future work \cite{toappear}.

We compute the indices for some values of $n$, $N_c$ and $N_f$. Results are listed in the table below.

\begin{longtable}{|c|c|c|p{7.5cm}|}
\hline
  &  Electric  &  Magnetic  &  \\
$(n,N_c,N_f)$  &  $SU(N_c)$  &  $U(1)\times U(nN_f- N_c)$  &  Index ($r$ is the IR $R$-charge of $Q$.)\\
\hline$$
$(2,1,1)$       &   free   &  $U(1)\times U(1)$    &  $1+x^{2/3}-4 x^2-3 x^{8/3}+\left(3+3 x^{2/3}\right) x^{2 r}+\left(4+4 x^{2/3}\right) x^{3 r}+5 x^{4 r}+x^r \left(2+2 x^{2/3}-4 x^2\right)+x^{-r}
\left(-2 x^2-2 x^{8/3}\right)+\cdots$  \\
\hline
$(2,2,1)$       &   $SU(2)$   &  $U(1)\times U(0)$    &  $1+x^{2/3}+x^{4/3}-4 x^2+x^{4-12 r}+x^{\frac{10}{3}-10 r}+\left(1+4 x^{2/3}\right) x^{4 r}+x^{6 r}+x^{2 r} \left(1+4 x^{2/3}+x^{4/3}\right)+x^{-2
r} \left(x^{2/3}+2 x^{4/3}+x^2\right)+x^{-4 r} \left(x^{4/3}+2 x^2+x^{8/3}\right)+x^{-6 r} \left(x^2+2 x^{8/3}\right)+x^{-8 r} \left(x^{8/3}+2 x^{10/3}\right)+\cdots$  \\
\hline
$(2,2,2)$       &   $SU(2)$   &  $U(1)\times U(2)$    &  $1+x^{2/3}+x^{4/3}-16 x^2-31 x^{8/3}+x^{\frac{16}{3}-8 r}+20 x^{4 r}+x^{2 r} \left(6+16 x^{2/3}+6 x^{4/3}\right)+x^{-4 r} \left(x^{8/3}+2 x^{10/3}+x^4\right)+\cdots$  \\
\hline
$(2,3,2)$       &   $SU(3)$   &  $U(1)\times U(1)$    &   $1+x^{2/3}+10 x^{4/3}+20 x^2+x^{4-12 r}+4 x^{\frac{8}{3}-6 r}+\left(10+26 x^{2/3}\right) x^{4 r}+20 x^{6 r}+4 x^{\frac{2}{3}+3 r}+x^{2 r} \left(4+8
x^{2/3}+24 x^{4/3}\right)+x^{-2 r} \left(4 x^{4/3}+12 x^2\right)+x^{-4 r} \left(x^{4/3}+3 x^2+12 x^{8/3}\right)+x^{-8 r} \left(x^{8/3}+3 x^{10/3}\right)+\cdots$ \\
\hline
$(3,2,1)$       &   $SU(2)$   &  $U(1)\times U(1)$    &   $1+\sqrt{x}+2 x+x^{3/2}-3 x^2+x^{4-8 r}+x^{4 r}+x^{2 r} \left(1+4 \sqrt{x}+5 x\right)+x^{-2 r} \left(x+2 x^{3/2}+3 x^2+2 x^{5/2}\right)+x^{-4 r}
\left(x^2+2 x^{5/2}+3 x^3\right)+x^{-6 r} \left(x^3+2 x^{7/2}\right)+\cdots$ \\
\hline
$(3,4,2)$       &   $SU(4)$   &  $U(1)\times U(2)$    &   $1+\sqrt{x}+12 x+47 x^{3/2}+154 x^2+x^{4-16 r}+\left(1+3 \sqrt{x}\right) x^{3-12 r}+4 x^{3-10 r}+\left(4+16 \sqrt{x}\right) x^{2-6 r}+\left(20+60
\sqrt{x}\right) x^{6 r}+35 x^{8 r}+x^{4 r} \left(10+26 \sqrt{x}+99 x\right)+x^{2 r} \left(4+8 \sqrt{x}+\left(36+116 \sqrt{x}\right) x\right)+x^{-2
r} \left(\left(4+16 \sqrt{x}\right) x+60 x^2\right)+x^{-4 r} \left(\left(1+3 \sqrt{x}\right) x+\left(17+55 \sqrt{x}\right) x^2\right)+x^{-8 r} \left(\left(1+3
\sqrt{x}\right) x^2+19 x^3\right)\cdots$ \\
\hline
$(4,2,1)$       &   $SU(2)$   &  $U(1)\times U(2)$    &   $1+x^{2/5}+2 x^{4/5}+2 x^{6/5}+2 x^{8/5}-3 x^2+\left(1+4 x^{2/5}\right) x^{4 r}+x^{2 r} \left(1+4 x^{2/5}+5 x^{4/5}+8 x^{6/5}\right)+x^{-2 r} \left(x^{6/5}+2
x^{8/5}+3 x^2+4 x^{12/5}\right)+x^{-4 r} \left(x^{12/5}+2 x^{14/5}\right)+\cdots$ \\
\hline
$(4,3,1)$       &   $SU(2)$   &  $U(1)\times U(1)$    &   $1+2 x^{2/5}+6 x^{4/5}+x^{2-10 r}+x^{\frac{8}{5}-8 r}+\left(1+3 x^{2/5}\right) x^{2 r}+x^{4 r}+x^{-4 r} \left(x^{4/5}+4 x^{6/5}\right)+x^{-2 r}
\left(x^{2/5}+4 x^{4/5}+10 x^{6/5}\right)+x^{-6 r} \left(x^{6/5}+4 x^{8/5}\right)+\cdots$ \\
\hline
\caption{Indices of $SU(N_c)$ gauge theory with adjoint matter and its dual theory.}\label{index}
\end{longtable}

Let's work out the operator contents for the few lower orders of the index results.
Consider $n=N_c=N_f=2$ case. The $x^{2/3}$ comes from  $X$ of the electric side and $\tilde X$ of the magnetic theory. In the   electric theory, $4x^{2r}$ term comes from the meson contribution $2x^{2r}$ and from baryon contribution $Q_1Q_2$, $\tilde Q_1\tilde Q_2$. On the dual side the singlet $M_0$ gives $4x^{2r}$ and $u_-q^1q^2$ and $u_+\tilde q^1\tilde q^2$ gives $2x^{2r}$.

We check another term $16x^{2r+2/3}$. On the electric side, $\mathrm{Tr}XQ\tilde Q$ counts  four, $QX\tilde Q$ counts four.
Another four come from $Q^a(XQ^b)$ and one from $QQ\mathrm{Tr}X$ but  they are not completely independent of each other.
 They satisfies $Q^1(XQ^2)-Q^2(XQ^1)-Q^1Q^2\mathrm{Tr}X=0$. So the baryon like operators give  4, and the similar anti-baryons
 operators (e.g.,$\tilde{Q}^a(X\tilde{Q}^b)$) give 4, summing to $16x^{2r+2/3}$. On the dual magnetic theory, the counting is
 basically the same as the electric case, except the mesons are mapped to singlet $M_j$ and baryons $B'$ and $\tilde B'$ are mapped
 to $b'$, $\tilde b'$.
From $Y$, there are $x^{8/3-4r}$ and the same term comes from $v_{0,+}v_{0,-}$ of magnetic theory.
There are also $x^{4-4r}$ which comes from $v_{1,+}v_{1,-}$.

We consider one more example. For the case of $N_c=N_f=1$ and $n=2$, the electric theory is a free theory with matter fields $Q$, $\tilde Q$ and $X$ and the dual magnetic theory has gauge group $U(1)\times U(1)$. The chiral ring elements of the electric theory is
simply the free chiral matter fields $Q$, $\tilde Q$ and $X$. We can read off it from the index expression. For example,  adjoint field $X$ gives $x^{2/3}$. Each $Q$ and $\tilde Q$ gives $x^r$, leading to $2x^r$ in the index. The term $3x^{2r}$ comes from $QQ,\tilde Q\tilde Q, Q\tilde Q$, the term $3x^{2r+2/3}$ comes from $XQQ, X\tilde Q\tilde Q, X Q\tilde Q$ and so on.

 On the dual magnetic theory, the adjoint matter field $\tilde X$ correspond to the $X$ of electric theory and it gives $x^{2/3}$ as expected. The monopole operator $u_\pm$ of the gauged $U(1)$ coupled with the matter fields $q$, $\tilde q$, so $u_+\tilde q$ and $u_-q$ correspond to the chiral fields $Q,\tilde Q$. We also have a singlet field $M_0$,  $(u_+\tilde q)^2$, $(u_- q)^2$ giving $3x^{2r}$ which corresponds to $ Q\tilde Q, \tilde Q\tilde Q, QQ$ of  the electric theory. The index doesn't have the term $x^{2-2r}$ which may correspond to $v_{0,+}v_{0,-}$ becaue the superpotential $W=M_0\tilde X q\tilde q+M_1q\tilde q +M_0v_{0,+}v_{0,-}$ makes it Q-exact. The last term of the superpotential or similar terms can be generated for special values
 of $(n, N_c,N_f)$\cite{jparkadjoint}\\
\indent Let's work out one more example where nonperturbative truncation of the chiral ring occurs.
 For terms in $x^{2r+2/3}$, naively there are four chiral ring element in the
 magnetic theory which are $M_1$, $\tilde XM_0$, $\tilde X (u_+ \tilde q)^2$ and $\tilde X (u_-q)^2$. The existence of
 $M_1$ could be a problem here because the electric theory is a free theory, the corresponding operator $QX\tilde Q$ is
  not different from $Q\tilde QX=XQ\tilde Q$. This means that $M_1$ should not be a chiral ring element and canceled by
   some other terms. If we consider R-charges of the possible states, there are four candidates canceling $M_1$, $(\tilde{v}_{0,+}\tilde{v}_{0,-}, \psi_{v_{1,+}}\tilde{v}_{0,+},\psi_{v_{1,-}}\tilde{v}_{0,-}, \psi_{v_{1,+}} \psi_{v_{1,-}})$. But normally they are canceled against each other due to the superpotential $v_{1,\pm}\tilde{v}_{0,\mp}$. But the state $\tilde{v}_{0,+}\tilde{v}_{0,-}$  does not exist for the case of $U(1)$ gauge theory because $\tilde{v}_{0,+}$, $\tilde{v}_{0,-}$ have opposite charges for the same Cartan sector, so they are paired up. Rest of the candidate states have the same energy states,
   two of them fermionic and one of them bosonic. Thus the $M_1$  truncated nonperturbatively. Therefore we have the right value $3x^{2r}$ from $\tilde XM_0$, $\tilde X (u_+ \tilde q)^2$ and $\tilde X (u_-q)^2$.\\
\\

\vskip 0.5cm  \hspace*{-0.8cm} {\bf\large Acknowledgements} \vskip
0.2cm

\hspace*{-0.75cm}  We thank Dongmin Gang, Eunkyung Koh for the collaboration at the initial stage of the project
and thank for Chiung Hwang, Hyungchul Kim for the discussions. We also thank Ofer Aharony, Shlomo S. Razamat, Nathan Seiberg and
Brian Willett for the correspondences.
J.P. was supported by the National Research Foundation of Korea
(NRF) grant funded by the Korea government (MEST) with the Grants No. 2012-009117,
2012-046278 and 2005-0049409 through the Center for Quantum Spacetime (CQUeST) of
Sogang University. JP  also appreciates APCTP for its stimulating environment for research.


\newpage




\begin{thebibliography}{1000}





  \bibitem{Giveon09}
  A. Giveon and D. Kutasov,
  ``Seiberg Duality in Chern-Simons Theory,''
  Nucl. Phys. {\bf B812} (2009) 1,
{\tt [arXiv:0808.0360 [hep-th]]}.

\bibitem{Niarchos08}
V. Niarchos, ``Seiberg Duality in Chern-Simons Theories with Fundamental and Adjoint Matter,''
 JHEP {\bf 0811} (2008) 001,
{\tt [arXiv:0808.2771 [hep-th]]}.

\bibitem{Niarchos} V. Niarchos, ``R-charges, Chiral Rings and RG Flows in Supersymmetric
Chern-Simons-Matter Theories,''
JHEP {\bf 0905} (2009) 054.

\bibitem{Kapustin10-2}
 A. Kapustin, B. Willett, I. Yaakov,
 ``Nonperturbative Tests of Three-Dimensional Dualities,''
{\tt [arXiv:1003.5694 [hep-th]]}.

\bibitem{Kapustin10}
A. Kapustin, B. Willett, I. Yaakov,
 ``Tests of Seiberg-like Duality in Three Dimensions,''
 {\tt [arXiv:1012.4021 [hep-th]]}.


 \bibitem{BashkirovKapustin11}
 D.~Bashkirov, A.~Kapustin,
``Dualities between N = 8 superconformal field theories in three dimensions,''
  JHEP {\bf 1105}, 074 (2011).
  [arXiv:1103.3548 [hep-th]].




 \bibitem{Spiridonov11}
 C. Krattenthaler, V. P. Spiridonov, G. S. Vartanov,
 ``Superconformal indices of three-dimensional theories related by mirror symmetry,''
 JHEP {\bf 1106} (2011) 008,
{\tt [arXiv:1103.4075 [hep-th]]}.

 \bibitem{JafferisYin11}
 D. Jafferis and X. Yin, ``{Duality
Appetizer},''  {\tt [arXiv:1103.5700 [hep-th]]}.


\bibitem{Kapustin11}
  A. Kapustin,
  ``Seiberg-like duality in three dimensions for orthogonal gauge
  groups, ''
  {\tt [arXiv:1104.0466 [hep-th]]}.

\bibitem{Willett11}
  B. Willett and I. Yaakov,
  ``N=2 Dualities and Z Extremization in Three Dimensions,''
  {\tt [arXiv:1104.0487 [hep-th]]}.

\bibitem{Kapustin11-2}
A. Kapustin, B. Willett, ``Generalized Superconformal Index for Three Dimensional Field Theories,''
 {\tt [arXiv:1106.2484 [hep-th]]}.


  \bibitem{Bashkirov1106}
 D. Bashkirov, ``Aharony duality and monopole operators in three
dimensions,'' {\tt [arXiv:1106.4110 [hep-th]]}.



\bibitem{HKPP11}
  C. Hwang, H. Kim, K.-J. Park, J. Park,
  ``Index computation for 3d Chern-Simons matter theory: test of Seiberg-like duality,''
  {\tt [arXiv:1107.4942 [hep-th]]}.

\bibitem{Gang11}
 D. Gang, E. Koh, K. Lee, J. Park,
 ``ABCD of 3d ${\cal N}=8$ and 4 Superconformal Field Theories,''
{\tt [arXiv:1108.3647 [hep-th]]}.

\bibitem{Berenstein11}
 D. Berenstein, M. Romo, ``Monopole operators, moduli spaces and dualities,''
{\tt [arXiv:1108.4013 [hep-th]]}.

\bibitem{Niarchos11}
 T. Morita, V. Niarchos, ``F-theorem, duality and SUSY breaking in one-adjoint Chern-Simons-Matter theories,''
 {\tt [arXiv:1108.4963 [hep-th]]}.


\bibitem{Benini11}
F. Benini, C. Closset, S. Cremonesi,
 ''Comments on 3d Seiberg-like dualities,''
 {\tt [arXiv:1108.5373 [hep-th]]}.





\bibitem{HPP11}
 C. Hwang, K. -J. Park, J. Park,
 ``Evidence for Aharony duality for orthogonal gauge groups,''
{\tt [arXiv:1109.2828 [hep-th]]}.


\bibitem{AharonyShamir11}
 O. Aharony, I. Shamir,
 ``On $O(N_c)$ d=3 N=2 supersymmetric QCD Theories,''
{\tt [arXiv:1109.5081 [hep-th]]}.


\bibitem{KapustinPark}
  A.~Kapustin, H.~Kim and J.~Park,
  ``Dualities for 3d Theories with Tensor Matter,''
  JHEP {\bf 1112}, 087 (2011)
  {\tt [arXiv:1110.2547 [hep-th]]}.

\bibitem{SeibergCS}

Authors: K. Intriligator, N. Seiberg,
``Aspects of 3d N=2 Chern-Simons-Matter Theories,''
{\tt [arXiv:1305.1633 [hep-th]]}.



 \bibitem{Jafferis10}
  D. Jafferis,
  ``The Exact Superconformal R-Symmetry Extremizes Z,''
{\tt [arXiv:1012.3210 [hep-th]]}.

\bibitem{Aharony97}
 O. Aharony ``IR duality in d=3 $N$=2
  $Usp(2N_c)$ and $U(N_c)$ Gauge Theories''
 Nucl. Phys.{\bf  B404} (1997) 71,
 {\tt [arXiv:hep-th/9703215]}.

\bibitem{Seibergsu}
 O. Aharony, S. Razamat, N. Seiberg, B. Willett
``3d dualities from 4d dualities,''
{\tt arXiv:1305.3924 [hep-th]}.

\bibitem{Seiberg3d}

 O. Aharony, A. Hanany, K. Intriligator, N. Seiberg, M.J. Strassler,
``Aspects of N=2 Supersymmetric Gauge Theories in Three Dimensions,''
 Nucl.Phys.{\bf B499} (1997) 67
{\tt [arXiv:hep-th/9703110]}

\bibitem{Witten-SL2}
 E. Witten, ``SL(2,Z) Action On Three-Dimensional Conformal Field Theories With Abelian Symmetry,''
{\tt [arXiv:hep-th/0307041]}.



 \bibitem{DGG}

 T. Dimofte, D. Gaiotto, S. Gukov
``Gauge Theories Labelled by Three-Manifolds,''
{\tt arXiv:1108.4389 [hep-th]}.

 \bibitem{DGG2}
  T.~Dimofte, D.~Gaiotto and S.~Gukov,
  ``3-Manifolds and 3d Indices,''
  {\tt arXiv:1112.5179 [hep-th]}.

\bibitem{jparkadjoint}
 H. Kim, J.  Park,
``Aharony Dualities for 3d Theories with Adjoint Matter,''
{\tt arXiv:1302.3645 [hep-th]}.

\bibitem{ours}
 D. Gang, E. Koh, S. Lee, J. Park,
``Superconformal Index and 3d-3d Correspondence for Mapping Cylinder/Torus,''
{\tt arXiv:1305.0937 [hep-th]}.

\bibitem{toappear} H. Kim, J. Park, K. Park, work on progress











\bibitem{Bhattacharya09}
  J. Bhattacharya and S. Minwalla,
  ``Superconformal Indices for $ N=6$ Chern Simons Theories,''
  JHEP, {\bf 0901} 2009) 014,
 {\tt [arXiv:0806.3251 [hep-th]]}.

\bibitem{Kim09}
  S. Kim,
  ``The complete superconformal index for N=6 Chern-Simons theory,''
  Nucl. Phys. {\bf B821} (2009) 241,
  {\tt [arXiv:0903.4172 [hep-th]]}.


\bibitem{Imamura11}
  Y. Imamura and S. Yokoyama,
  ``Index for three dimensional superconformal field theories with general R-charge assignments,''
  {\tt [arXiv:1101.0557 [hep-th]]}.









\end{thebibliography}
\end{document}